\documentclass[aps,prd,twocolumn,superscriptaddress,groupedaddress,nofootinbib]{revtex4-1}

\usepackage{amsmath,amssymb,bm,color,comment,dcolumn,mathrsfs,tabularx,graphicx}
\usepackage[hidelinks=true]{hyperref}
\hypersetup{colorlinks=true,linkcolor=blue,citecolor=blue,urlcolor=blue}
\usepackage{Macro}

\def\lmax{\l_{\rm max}}

\begin{document}

\title{The CMB lensing bi-spectrum as a probe of modified gravity theories}

\author{Toshiya Namikawa}
\affiliation{Leung Center for Cosmology and Particle Astrophysics, National Taiwan University, Taipei, 10617, Taiwan}

\author{Fran\c{c}ois R. Bouchet}
\affiliation{Institut d'Astrophysique de Paris, UMR7095, CNRS \& Sorbonne Universit{\'e}, Paris, France}
\affiliation{Yukawa Institute for Theoretical Physics, Kyoto University, Kyoto 606-8502, Japan}

\author{Atsushi Taruya}
\affiliation{Center for Gravitational Physics, Yukawa Institute for Theoretical Physics, Kyoto University, Kyoto 606-8502, Japan}
\affiliation{Kavli Institute for the Physics and Mathematics of the Universe (WPI), The University of Tokyo Institutes for Advanced Study, The University of Tokyo, 5-1-5 Kashiwanoha, Kashiwa, Chiba 277-8583, Japan}

\date{\today}

\begin{abstract}
Cosmological structures grow differently in theories of gravity which are modified as compared to Einstein's General relativity (GR). Cosmic microwave background (CMB) fluctuation patterns at the last scattering surface are lensed by these structures along the photon path to the observer. The observed CMB pattern therefore keeps trace of the growth history of structures. 
We show that observations of the CMB lensing bi-spectrum offer an interesting way to constrain deviations from GR in a broad class of scalar-tensor theories of gravity called ``beyond Horndeski". 
We quantify how the constraints on generic parameters describing the deviations from GR depend on the effective multipole range of the analysis. Our results further indicate that an accurate nonlinear correction of the matter bi-spectrum in the modified gravity considered is necessary when the bi-spectrum is used to probe scales beyond a multipole $\lmax\agt 1500$. We also found that the results are insensitive to details of the implementation of the screening mechanism, at very small scales. 
We finally demonstrate the potential of the lensing bi-spectrum to provide a blind reconstruction of the redshift evolution of our modified gravity parameters by combining the analysis of CMB and low-z source lensing data.
\end{abstract}

\maketitle


\section{Introduction} \label{sec:intro}

Dark energy is an enduring mystery. One possibility is that the observational evidence pointing to its existence, i.e., late-time cosmic acceleration, is misinterpreted and should rather be considered as traces of the breakdown of General relativity (GR) at large scales. Many experiments are being conducted in order to test this hypothesis, in particular by seeking a possible deviation from GR in the growth rate of large scale structures of the Universe (e.g., \cite{BOSS:overview, eBOSS, DESI, DES, Euclid12}). 

One fundamental difficulty in devising tests of gravity on cosmological scales is that the large scale structures of the Universe are delineated by visible objects -- galaxies, clusters of galaxies -- which are biased tracers of the dark matter distribution, while this is the latter which is the dominant source of gravity at these scales. Of course, there are well developed models relating the light distribution to the mass distribution and a perturbative description of the galaxy bias should  work well at large scales, only introducing a small number of nuisance parameters (e.g., \cite{Desjacques:2016}). 
Nevertheless, deviations from GR are bound to be very small, and controlling the accuracy of these models at the required level of precision is a daunting challenge which will be hard to meet convincingly. It is therefore all too natural to try finding observables which rely on a solid understanding of their physics. Gravitational lensing effects are obvious candidates, which are very promising. However, practical analyses of the weak lensing probed with luminous objects, referred to as cosmic shear, still have to cope with a number of difficulties, e.g., the alignment of objects with their surrounding structures or uncertainties in the redshift distribution of the lensed sources. In addition, most of the lensing effect comes from redshifts around half that of the most distant sources, typically $z\sim1$. One interesting avenue to explore will be the lensing effect on the Lyman-$\alpha$ forest on the line of sight of distant quasars (e.g., \cite{1996ApJ...457..529B, 2007JCAP...11..008L, 2018MNRAS.477.2841M}). Here, we instead look at the lensing effect on the CMB, whose physical origin is very well known. In that case, the main contribution to the lensing signal comes from a large redshift range centered at $z\sim 2$, and this allows us to probe a redshift range hardly attainable by the lensing effects probed with luminous sources. 

Multiple works have already considered the future forecast to constrain modified gravity theories from the CMB lensing measurements, focusing in particular on the (angular) power spectrum and/or cross-power spectra with other cosmological observables(e.g., \cite{Acquaviva:2005xz, Calabrese:2009, Amendola:2014wma, Munshi:2014tua, DAmico:2017}). But these two-point statistics are far from being exhaustive, and further information can be extracted with the higher-order statistics. Refs.~\cite{Namikawa:2016b, Pratten:2016} showed that the lensing bi-spectrum of the CMB is detectable at high statistical significance in near-term experiments. It is thus natural, as a second step, to explore the potential power of measurements of the CMB lensing bi-spectrum in assessing modified gravity theories. The aim of this paper is to present the first forecast study of this probe.

This paper is organized as follows. 
Section~\ref{bisp} recalls the basic theory of the lensing bi-spectrum and how it is altered in modified gravity theories. 
Section~\ref{results} presents our main results, while
Section~\ref{summary} summarizes our work and discusses perspectives. 

Throughout the paper, we adopt a fiducial cosmological model which is consistent with the latest Planck cosmology \cite{P15:main}, i.e., a spatially flat $\Lambda$CDM cosmology with 
the baryon and matter density, $\Omega_{\rm b}h^2=0.0223$ and $\Omega_{\rm m}h^2=0.119$, 
the dark-energy density $\Omega_\Lambda=0.689$, the amplitude of the primordial scalar 
power spectrum, $A_{\rm s}=2.13\times 10^{-9}$, and its spectral index at 
$k=0.05\mathrm{Mpc}^{-1}$, $n_{\rm s}=0.965$, with the reionization optical depth, $\tau=0.0630$. 

\section{Lensing Bi-spectrum} \label{bisp}

In this section, we first review briefly the formalism of the CMB lensing bi-spectrum in GR 
(see, e.g., \cite{Takada:2004, Namikawa:2016b}). 
We then describe the modification of the bi-spectrum in a general class of modified gravity theories (see, e.g., \cite{Takushima:2014, Yamauchi:2017, Hirano:2018}). 

\subsection{Lensing potential} 

The gravitational lensing effect on the CMB anisotropies is described as a remapping 
of the CMB fluctuations at the recombination by the so-called deflection angle, $\bm{d}=\bn\phi$, 
\footnote{
Here we ignore the curl mode of the deflection angle (e.g., \cite{Namikawa:2011b}) because 
it contribution is negligible compared to that of the lensing potential in 
the standard $\Lambda$CDM cosmology (e.g., \cite{Saga:2015}). 
}
where $\phi$ is the lensing potential (e.g. \cite{Lewis:2006fu,Hanson:2009kr}):
\al{
	\phi(\hatn) = -2\INT{}{\chi}{}{0}{\chi_*} W(\chi,\chi_*) \Psi(\chi,\hatn)
    \,.
}
Here $\chi_*$ is the comoving distance to the CMB last-scattering surface, and $\Psi$ is the 
Weyl potential. The lensing kernel, $W(\chi,\chi_*)$, is defined (for a flat cosmology) as
\al{
	W(\chi,\chi_*) = \frac{\chi_*-\chi}{\chi\chi_*}  \,. 
}
Let us define the lensing convergence as $\kappa=-\bn^2\phi/2$. By using Poisson equation, we obtain 
\al{
	\kappa(\hatn) = \INT{}{\chi}{}{0}{\chi_*} 
    	\frac{3\Omega_{\rm m,0}H_0^2\chi^2}{2a(\chi)}\,W(\chi,\chi_*) \delta_{\rm m}(\chi,\hatn)
    \,,
    \label{eq:kappa_delta}
}
where $a$ is the scale factor and $\delta_{\rm m}$ is the underlying density fluctuations of matter along the line-of-sight. 
The lensing potential (and thus the convergence) can be reconstructed from the observed 
CMB anisotropies by using the fact that a fixed lensing potential introduces statistical
anisotropy into the observed CMB (e.g., \cite{Hu:2001, Hirata:2002jy, 2006PhR...429....1L}). 
The reconstructed lensing convergence map can then be used for cosmology by first 
transforming the map into its harmonic coefficients, $\kappa_{\l m}$, and then by
measuring various moments like the power spectrum and the bi-spectrum (which entails an accurate subtraction of the non-lensing contributions, see e.g., \cite{Hanson:2010rp, Namikawa:2012}). 
Recent CMB experiments have already detected the power spectrum of the lensing potential very precisely \cite{ACT16:phi, BKVIII, P15:phi, PB14:phi, SPT:phi}. For instance, the {\it Planck} detection \citep{P15:phi} has a 40\,$\sigma$ significance. The detection and precise determination of the CMB lensing bi-spectrum are therefore obvious and important next step in CMB scientific analyses. 

\subsection{Lensing bi-spectrum}
\label{subsec:lensing_bispec}

The bi-spectrum of the lensing convergence, defined in harmonic space, has translational and rotational invariance. This is true as long as the statistical isotropy holds. It is thus sufficient  to characterize it with a function $B_{\l_1\l_2\l_3}$ of only three variables, weighted by the Wigner $3j$-symbols through
\al{
	\ave{\kappa_{\l_1m_1}\kappa_{\l_2m_2}\kappa_{\l_3m_3}} 
		= \Wjm{\l_1}{\l_2}{\l_3}{m_1}{m_2}{m_3}B_{\l_1\l_2\l_3}
	\,.
}
In what follows, we use the flat-sky approximation, and start by computing the bi-spectrum given by 
\al{
	\ave{\kappa_{\bl_1}\kappa_{\bl_2}\kappa_{\bl_3}} 
		= (2\pi)^2\delta(\bl_1+\bl_2+\bl_3)\,B_{\bl_1\bl_2\bl_3}
	\,.
}
The full-sky bi-spectrum is then obtained from the flat-sky bi-spectrum through
\al{
	B_{\l_1\l_2\l_3} & =  \Wjm{\l_1}{\l_2}{\l_3}{0}{0}{0}
		\sqrt{\frac{(2\l_1+1)(2\l_2+1)(2\l_3+1)}{4\pi}} \nonumber \\
                     & \times B_{\bl_1\bl_2\bl_3}
	\,,
}
where the multipoles have to satisfy the triangle condition $|\l_i-\l_j|\leq\l_k\leq\l_i+\l_j$. 
In order to evaluate this expression, it is enough at the scales of interest to use the following approximate form of the Wigner $3j$-symbol,
\al{
	\Wjm{\l_1}{\l_2}{\l_3}{0}{0}{0} &\simeq (-1)^L\sqrt{\frac{\E^3}{2\pi}}(L+1)^{-1/4}
		\notag \\
		&\times\,\prod_{i=1}^3(L-\l_i+1)^{-1/4}\left(\frac{L-\l_i+1/2}{L-\l_i+1}\right)^{L-\l_i+1/4}
	\,
}
for even $L$, where we define $L=(\ell_1+\ell_2+\ell_3)/2$. For odd $L$, the Wigner $3j$-symbol becomes zero. 

The CMB lensing bi-spectrum is sourced by \begin{enumerate}
\item the nonlinear evolution of the large-scale structure \cite{Namikawa:2016b}, and 
\item the so-called post-Born correction \cite{Pratten:2016}, i.e., the correction to the Born approximation for which the lensing effect is evaluated on the unperturbed geodesic.
\end{enumerate}
We denote these contributions respectively by $B^{\rm LSS}$ and $B^{\rm pb}$, and present their explicit expressions below.  

The CMB lensing bi-spectrum from the nonlinear growth of the density perturbations is given in the flat-sky limit by \cite{Namikawa:2016b}
\al{
	B^{\rm LSS}(\bl_1,\bl_2,\bl_3) &= \INT{}{\chi}{}{0}{\chi_*} 
		\left[\frac{3\Omega_{m,0}H_0^2}{2a(\chi)}\right]^3
		\notag \\
		&\times \chi^2 W^3(\chi,\chi_*)\, B_\delta(\bk_1,\bk_2,\bk_3,\chi)
	\,. \label{Eq:bisp:lss}
}
Here, $B_\delta$ is the \emph{matter} bi-spectrum arising from the nonlinear growth of structure. In the weakly nonlinear regime, it can be obtained by using perturbation theory. The result at the tree-level order is of the general form
\al{
	B_\delta(\bk_1,\bk_2,\bk_3,\chi) 
		&= 2F_2(\bk_1,\bk_2,z)\,P_{\delta}(k_1,z)P_{\delta}(k_2,z) 
	\notag \\
		&+ 2\,{\rm perms.}
	\, ,
}
where $P_\delta(k,z)$ is the matter power spectrum at redshift $z(\chi)$, and the function $F_2$ is the second-order perturbation theory kernel (e.g., \cite{Bernardeau:2001qr}). 
Writing $\bk_1\cdot\bk_2=k_1k_2\cos\theta$, it is given by
\al{
	F_2(\bk_1,\bk_2,z) &= \frac{5}{7}\,a(k_1,z)a(k_2,z) 
		\notag \\
		&+ \frac{1}{2}\,\frac{k_1^2+k_2^2}{k_1k_2}\,b(k_1,z)b(k_2,z)\cos\theta 
		\notag \\
		&+ \frac{2}{7}\,c(k_1,z)c(k_2,z)\cos^2\theta 
	\,, \label{Eq:F2:GR}
}
where $a(k,z)$, $b(k,z)$ and $c(k,z)$ are unity at the tree-level of perturbation theory. 
In the highly nonlinear regime, the deviation from tree-level prediction are significant, and a proper treatment of the nonlinear effects coming from the higher-order perturbations is needed. The scale- and time-dependent coefficients $a$, $b$, and $c$ effectively characterize these, and their deviation from unity is calibrated with high-resolution $N$-body simulations. 
According to \cite{Scoccimarro:2001}, they are given by
\al{
	a(k,z) &= \frac{1+\{\sigma_8(z)\}^{a_6}\sqrt{0.7Q(n_{\rm eff})}(qa_1)^{n_{\rm eff}+a_2}}{1+(qa_1)^{n_{\rm eff}+a_2}} \\
	b(k,z) &= \frac{1+0.2a_3(n_{\rm eff}+3)(qa_7)^{n_{\rm eff}+3+a_8}}{1+(qa_7)^{n_{\rm eff}+3.5+a_8}} \\
	c(k,z) &= \frac{1+[4.5a_4/(1.5+(n_{\rm eff}+3)^4)](qa_5)^{n_{\rm eff}+3+a_9}}{1+(qa_5)^{n_{\rm eff}+3.5+a_9}}
	\,, 
}
with $Q(x)=(4-2^x)/(1+2^{x+1})$. Here, the variable $q$ is given by $q=k/k_{\rm NL}$ with 
the nonlinear scale, $k_{\rm NL}$, determined by $4\pi k^3_{\rm NL}P_{\rm m}^{\rm lin}(k_{\rm NL})=1$. 
The quantity $\sigma_8(z)$ is the variance of the matter density 
fluctuations smoothed with a top-hat sphere of radius $8h^{-1}$Mpc at redshift $z$. 
The logarithmic slope, $n_{\rm eff}\equiv d\ln P_{\rm m}^{\rm lin}(k)/d\ln k$, 
is the effective spectral index of the linear power spectrum, $P_{\rm m}^{\rm lin}(k)$.  
The parameters, $a_i$, are determined by fitting results of N-body simulations, which yields \cite{Scoccimarro:2001}
\al{
	a_1 &= 0.250 & a_2 &= 3.50 & a_3 &= 2.00 \notag \\
    a_4 &= 1.00 & a_5 &=2.00 & a_6 &= -0.200 \notag \\
    a_7 &= 1.00 & a_8 &= 0.00 & a_9 &= 0.00 \,.
    \label{Eq:SC}
}
Later on, Ref.~\cite{Gil-Marin:2012} proposed an improved fit given by  
\al{
	a_1 &= 0.484 & a_2 &= 3.74 & a_3 &= -0.849 \notag \\
    a_4 &= 0.392 & a_5 &= 1.01 & a_6 &= -0.575 \notag \\
    a_7 &= 0.128 & a_8 &= -0.722 & a_9 &= -0.926 \,.
    \label{Eq:GM}
}
In our baseline calculations, we use the parameters of Ref.~\cite{Gil-Marin:2012} (hereafter ``GM")  
but we shall also use the earlier results of Ref.~\cite{Scoccimarro:2001} (hereafter ``SC")
as a means to assess the dependence of our results on the accuracy of these fitting formula.

The post-Born correction to the CMB lensing bi-spectrum, $B^{\rm pb}$ is given as \cite{Pratten:2016}
\al{
	B^{\rm pb}(\bl_1,\bl_2,\bl_3) 
		&= 2\frac{\bl_1\cdot\bl_2} {\l^2_1\l^2_2}         \notag \\
        &\times\left[\bl_1\cdot\bl_3\, M(\l_1,\l_2)+\bl_2\cdot\bl_3\,M(\l_2,\l_1)\right] 
		\notag \\
		&+ {\rm cyc. perm.} 
	\,, \label{Eq:bisp:pb}
}
where 
\al{
	M(\l,\l') &= \l^4\INT{}{\chi}{}{0}{\chi_*} \frac{[W(\chi,\chi_*)]^2}{\chi^2} 
		\notag \\
		&\times P_\Psi \left(\frac{\l}{\chi},\chi\right) C^{\kappa\kappa}_{\l'}(\chi,\chi_*)
	\,.
}
$P_\Psi(k,\chi)$ is the power spectrum of the Weyl potential at a comoving distance $\chi$, and
\al{
C_\l^{\kappa\kappa}(\chi',\chi_*) &= \l^4\INT{}{\chi}{}{0}{\chi'}\frac{W(\chi,\chi')W(\chi,\chi_*)}{\chi^2}
		\notag \\
		&\times P_\Psi\left(\frac{\l}{\chi},\chi\right) .
}
If these terms are known, or determined accurately, then the post-Born term is known as well.

All in all, the CMB lensing bi-spectrum is the sum of \eqs{Eq:bisp:lss,Eq:bisp:pb}. 

\subsection{Effect of modified gravity on the bi-spectrum}\label{bispec_models}

In modified gravity theories, the perturbation theory kernel $F_2$ is altered and may be written, at the tree level in the quasi static approximation, as
(e.g., \cite{Yamauchi:2017,Hirano:2018})
\al{
	F_2(\bk_1,\bk_2,z) &= \left(\kappa(z)-\frac{2}{7}\lambda(z)\right)
		\notag \\
		&+ \kappa(z)\frac{1}{2}\frac{k_1^2+k_2^2}{k_1k_2}\cos\theta 
		\notag \\
		&+ \lambda(z)\frac{2}{7}\cos^2\theta 
	\,. \label{Eq:MG:F2:tree}
}
The kernel above coincides with the GR case when $\lambda(z)=\kappa(z)=1$. Let us note that here and in all the following we use $\kappa$ as a parameter characterizing the deviation from GR, as in many previous theoretical papers. This is then not to be confused with the lensing convergence which is denoted by the same symbol in other parts of the literature (and in \eqref{eq:kappa_delta}). 

The Horndeski theory of gravity (e.g., Refs.~\cite{Horndeski1974,Kobayashi_etal2011,Deffayet2011}), is  the most general non-degenerate scalar-tensor theory in 4D space-time that leads to second-order equations of motion. It may have $\lambda\ne1$ in general, but $\kappa=1$ 
is still preserved. An even wider class of theories imaginatively called ``beyond Horndeski" theories, including GLPV \cite{Gleyzes_etal2015a, Gleyzes_etal2015b} and DHOST \cite{Langlois_Noui2016a, Langlois_Noui2016b}, can explicitly break this latter condition, in close connection with the violation of Vainshtein mechanism to recover GR at nonlinear regime. Testing and constraining possible deviations of $\lambda$ and $\kappa$ from unity is thus very interesting, and could give important information on gravity at cosmological scales, rather independently of the growth rate of structure probed with galaxy redshift surveys. Further, no strong constraint on $\lambda$ has been obtained so far, and no theoretical upper/lower limits is known for $\kappa$
(see, e.g., Refs.~\cite{Takushima:2014, Yamauchi:2017, Hirano:2018} for further discussion 
of the possible values of $\lambda$ and $\kappa$ under some specific models).
The measurement of the bi-spectrum is therefore key to narrow down the constraints.

A particular subclass of modified gravity theories may have a specific redshift dependence of $\lambda$ and $\kappa$, and such a form will have to be used ultimately to get the tightest constraints on these specific theories. Here, we rather wish to look at the \emph{generic} potential of the bi-spectrum probe. Following Ref.~\cite{Yamauchi:2017}, we adopt the functional form of $\lambda$ and $\kappa$ as
\al{
	\lambda(z) &= [\Omega_m(z)]^{\xi_\lambda} \,, \notag \\ 
	\kappa(z) &= [\Omega_m(z)]^{\xi_\kappa} \,, 
}
where $\Omega_m(z)=\Omega_{m,0}/(\Omega_{m,0}+a^3\Omega_\Lambda)$. This naturally embodies the expectation that a modified theory will converge to GR at high-$z$ and preserve the successful predictions of CMB anisotropies using GR. The form is also monotonic, in keeping with the idea of generic constraints. Our goal is therefore to assess how well the generic parameters $\xi_\lambda$ and $\xi_\kappa$ can be constrained by measurement of the CMB lensing bi-spectrum. 

One should recall that the modification of the kernel $F_2$ given in \eq{Eq:MG:F2:tree} is only valid in the weakly nonlinear regime. In order to improve the constraints, we may want to use measurements at small scales, taking into account the nonlinear corrections introduced earlier in Sec.~\ref{subsec:lensing_bispec}. This is, however, not trivial in the context of modified gravity, because a modification of gravity can also change the nonlinear corrections, and a proper account of these needs more elaborate work which we leave for future investigation. Here we rather adopt the fitting formula given in the GR case, in order to assess the potential power of the CMB lensing bi-spectrum in the intermediate regime. Therefore the kernel $F_2$ used in our analysis is given by
\al{
	F_2(\bk_1,\bk_2,z) &= \left(\kappa(z)-\frac{2}{7}\lambda(z)\right)a(k_1,z)a(k_2,z)
		\notag \\
		&+ \kappa(z)\frac{1}{2}\frac{k_1^2+k_2^2}{k_1k_2}\,b(k_1,z)b(k_2,z)\cos\theta 
		\notag \\
		&+ \lambda(z)\frac{2}{7}c(k_1,z)c(k_2,z)\cos^2\theta 
	\,. \label{Eq:MG:F2:loop}
}
We will then compare the forecast results based on Eq.~(\ref{Eq:MG:F2:loop}) with those derived from the tree-level kernel at Eq.~(\ref{Eq:MG:F2:tree}). The impact of the modification of this formula at small scales is also discussed in detail (see Sec.~\ref{subsec:screening}).

Let us conclude this theoretical section with a couple of comments. The first concerns the relation between the lensing potential and the density field, which was given in Eq.~(\ref{eq:kappa_delta}). This equation may be altered in modified gravity theories, and the expression given at Eq.~(\ref{Eq:bisp:lss}) might not be relevant. Fortunately, Ref.~\cite{DAmico:2017} showed that the modified Weyl potential is given by a simple scaling of the matter density fluctuations by a factor $\mu$. This means that what is actually constrained from the lensing bi-spectrum are combinations of the parameters, i.e., $\mu^3\lambda$ and $\mu^3\kappa$. This being understood, in the following, we keep using the same notation, and denote these scaled parameters simply by $\lambda$ and $\kappa$.

The second point concerns the post-Born correction term, which could also receive corrections from the modification of gravity. However, the impact of such a modification is expected to be very small because the measurement of the lensing power spectrum severely limits a modification to the post-Born correction in the bi-spectrum. Indeed, the signal-to-noise ratio of the lensing power spectrum will be very high in future experiments ($\sim \mC{O}(10^3)$) and the allowed modification to the power spectrum amplitude is therefore smaller than about $0.1$\%. The signal-to-noise ratio of the bi-spectrum is, on the other hand, much lower than that of the power spectrum ($\sim \mC{O}(10)$), which is equivalent to a $\sim 10$\% constraint on the modification of the bi-spectrum amplitude. This means that the prior information from the power spectrum limits the modification to the post-Born correction well below the measurement uncertainty of the bi-spectrum. For this reason, in our analysis, we ignore the effect of possible modification to the post-Born correction.

\section{Results} \label{results}

\begin{figure*}
\bc
\includegraphics[width=89mm,clip]{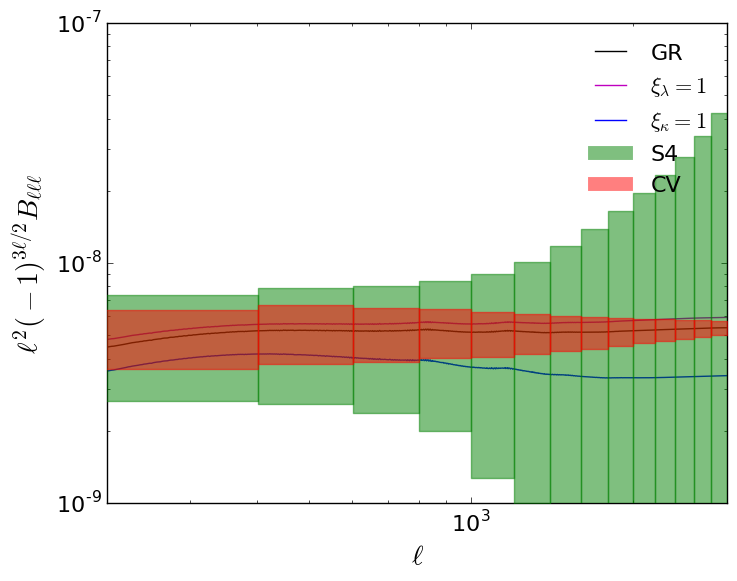}
\includegraphics[width=89mm,clip]{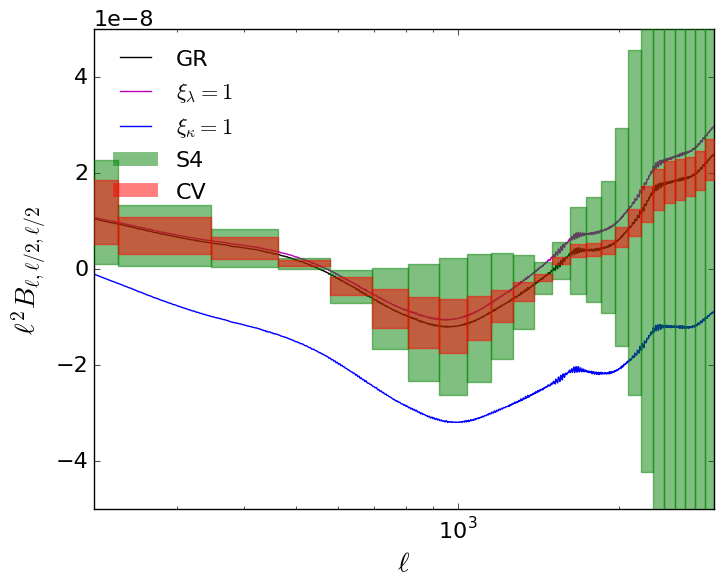}
\caption{The CMB lensing bi-spectrum in GR and modified theories of gravity ($\xi_\lambda=1$ or $\xi_\kappa=1$) with error bars expected from a future CMB experiment (S4) and in the cosmic-variance limit (CV), see specifications in Table~\ref{tab:specs}. 
The left and right panel show the case for an equilateral and folded configuration with, 
respectively, $\l_1=\l_2=\l_3\equiv\l$ and $\l_1=2\l_2=2\l_3\equiv\l$. 
}
\label{fig:bispec_equi_iso}
\ec
\end{figure*}

\begin{table}
\bc
\caption{
Specifications for the CMB experiments considered in this paper: the noise level in 
the CMB polarization map, $\Delta_{\rm P}$, in unit of $\mu$K\,arcmin, the angular resolution as expressed by the FWHM of a Gaussian beam, $\theta$, in unit of arcmin, and the fractional sky coverage, $f_{\rm sky}$.\label{tab:specs}
}
\label{Table:cmb} \vs{0.5}
\begin{tabular}{lccc} \hline\hline
        & $\Delta_{\rm P}$ [$\mu$K\,arcmin] & $\theta$ [arcmin] & $f_{\rm sky}$ \\ \hline 
CMB-S4  & 1 & 3 & 0.4  \\ 
CV      & 0 & --- & 1.0 \\ \hline
\end{tabular}
\ec
\end{table}

\begin{figure*}
\bc
\includegraphics[width=89mm,clip]{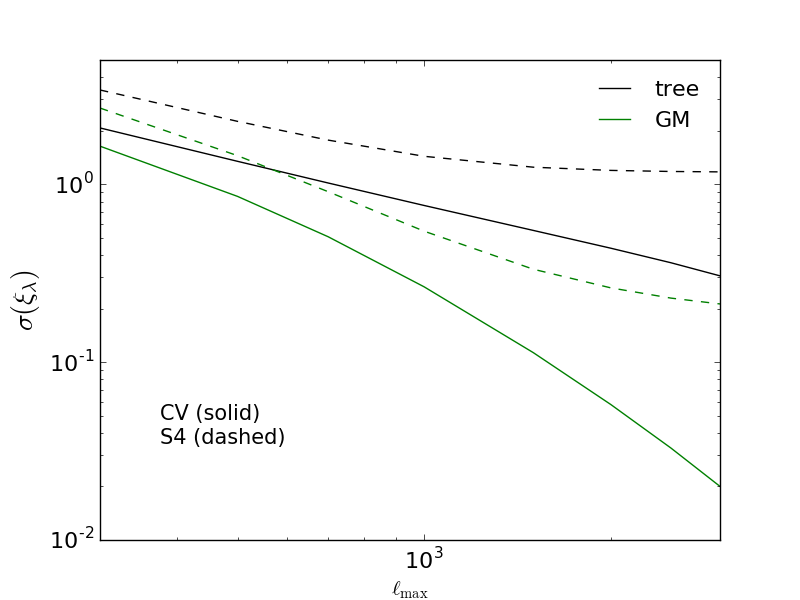}
\includegraphics[width=89mm,clip]{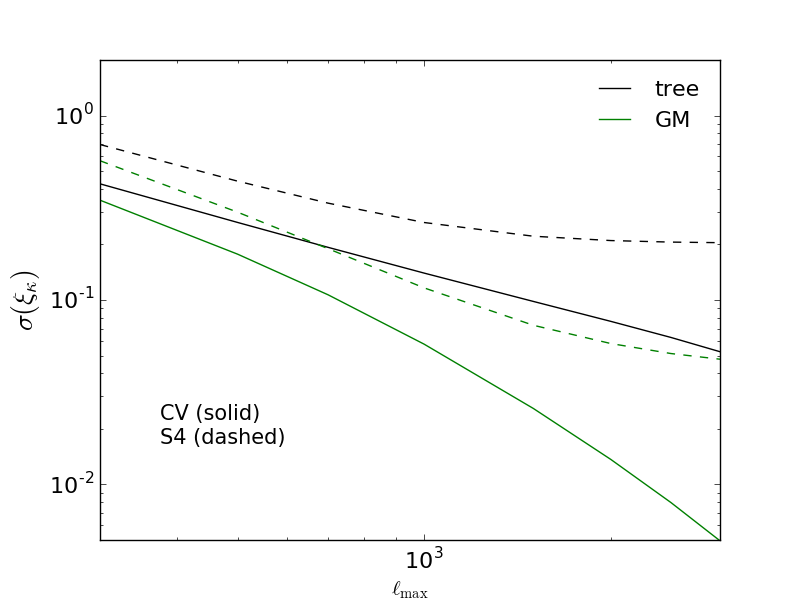}
\caption{
Left: Expected constraints on $\xi_\lambda$ as a function of $\lmax$ and experimental specifications, with and without the nonlinear correction (see text). The constraints are derived by marginalizing only $\xi_\lambda$ as a free parameter. Right: Same as Left but for $\xi_\kappa$. Note that the curves labeled ``tree" correspond to \eq{Eq:MG:F2:tree} and those labeled ``GM" to \eq{Eq:MG:F2:loop}.
}
\label{fig:const:usig}
\ec
\end{figure*}

\begin{figure*}
\bc
\includegraphics[width=89mm]{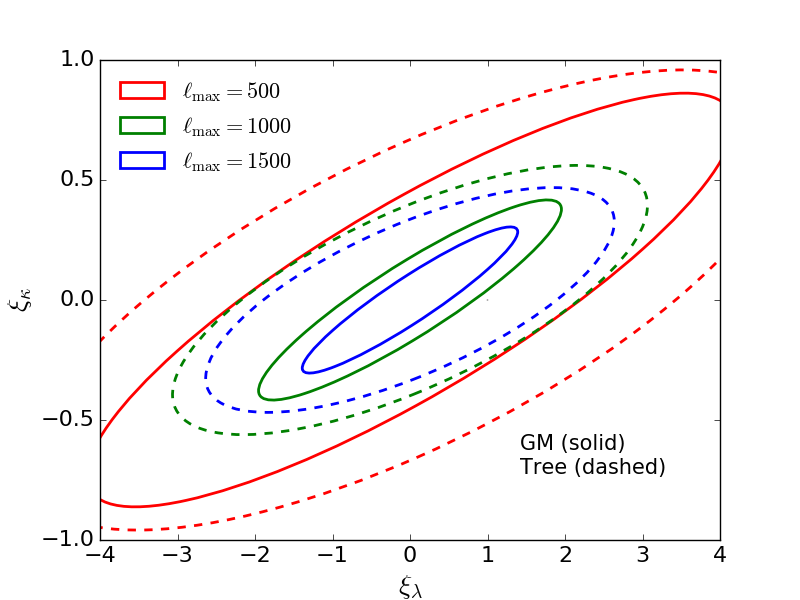}
\includegraphics[width=89mm]{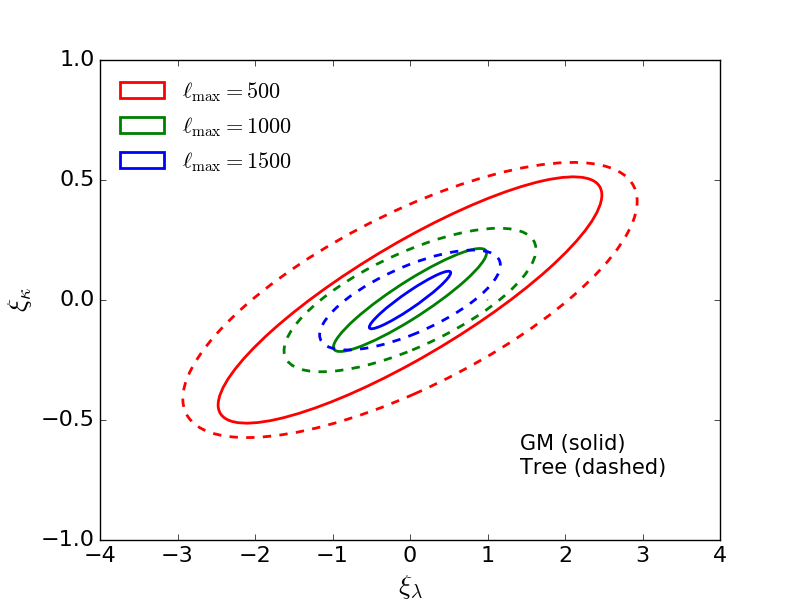}
\caption{
The expected joint constraints on the two parameters, $\xi_\lambda$ and $\xi_\kappa$, 
assuming the CMB-S4 experiment (Left) and in the cosmic-variance case (Right). 
Ellipses of different colors correspond to changes in the maximum multipole of the bi-spectrum used to constrain the parameters ($\lmax=500,1000,1500$). We also show with dashes the case when including a correction of the nonlinear evolution of the large-scale structure in the bi-spectrum (denoted by ``GM'').
}
\label{fig:const:2d}
\ec
\end{figure*}

\begin{figure*}
\bc
\includegraphics[width=89mm,clip]{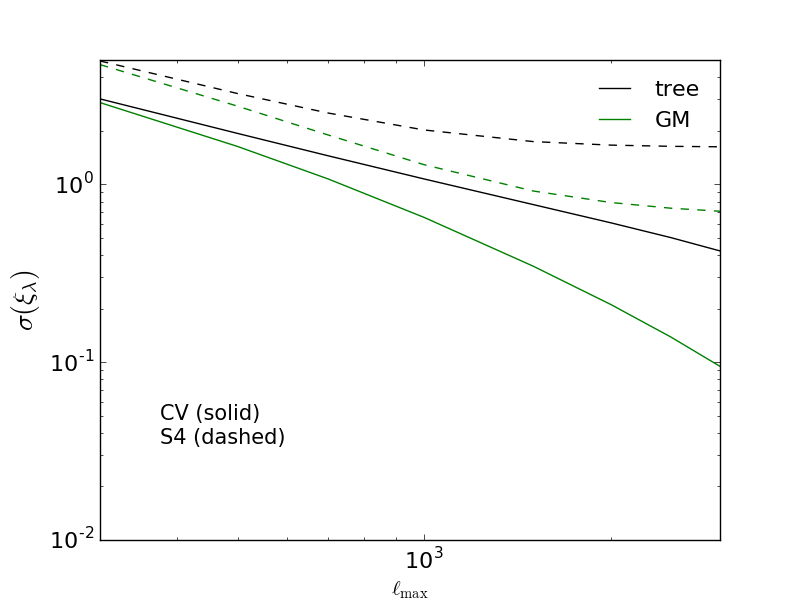}
\includegraphics[width=89mm,clip]{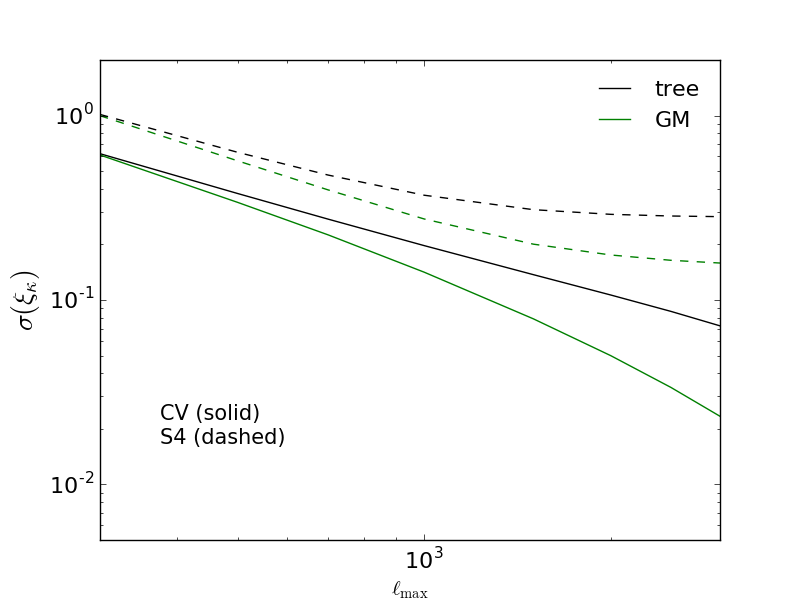}
\caption{
Expected constraints on $\xi_\lambda$ (Left) and $\xi_\kappa$ (Right) 
as a function of $\lmax$ with and without the fitting model of the 
nonlinear correction and experimental specifications. 
Compared to Fig.~\ref{fig:const:usig}, we assume that 
the two parameters are simultaneously constrained. 
}
\label{fig:const:sig}
\ec
\end{figure*}

\subsection{Expected constraints on parameters of modified gravity theories}

We begin by inspecting the expected signal for lensing bi-spectrum for various experiments and configurations. In Fig.~\ref{fig:bispec_equi_iso}, the bi-spectra for equilateral ($\l_1=\l_2=\l_3\equiv\l$, left) and folded ($\l_1=2\l_2=2\l_3\equiv\l$, right) configurations are shown, and the results are plotted as a function of $\l$.\footnote{
The error bars are computed as follows. 
We first choose the width of the multipole bin and 
bin centers of the multipole, $\ell_1$, $\ell_2$ and $\ell_3$. 
Using the three multipole bin centers, the binned equilateral and folded bi-spectra satisfy $\ell_1=\ell_2=\ell_3$ and $\ell_1=2\ell_2=2\ell_3$, respectively. 
For each binned bi-spectrum, $B_{\ell_1,\ell_2,\ell_3}$, 
its signal-to-noise (i.e. the inverse of the error on its amplitude) is computed. 
The error bars plotted in the figure are obtained by multiplying $B_{\ell_1,\ell_2,\ell_3}$ 
to the errors of the bi-spectrum amplitude at each bin. 
}
Solid lines are the expected bi-spectrum signals in GR (black) and modified gravity models ($\xi_\lambda=1$ for red, $\xi_\kappa=1$ for blue), which are computed following the prescription in Sec.~\ref{bispec_models}. The modification to gravity alters the scale-dependence of the bi-spectrum amplitude through the shape dependence of the kernel $F_2$, but only weakly, and the main effect appears to be a simple rescaling. 
In the folded bi-spectrum, however, the post-Born correction has non-negligible contributions to the bi-spectrum, and the total bi-spectrum has a bit complicated behavior.
In the figure, the statistical errors are depicted by shaded areas, both for the upcoming experiment, CMB-S4 (green), and an hypothetical full-sky experiment limited only by the cosmic variance (red). Note that the quoted error is estimated from the Gaussian noise used in the Fisher matrix (see below). 

Fig.~\ref{fig:bispec_equi_iso} implies that with a realistic measurement of the lensing bi-spectrum, one can simultaneously constrain both modified gravity parameters. We note that these configurations at large scales ($\ell \alt 1000$) are a simple rescaling of the GR case, and as such might be hard to disentangle from other observational effects, a further reason to hope for small scale measurements. In addition, the folded bi-spectrum is much more sensitive to the modified gravity parameters than the equilateral bi-spectrum because the amplitude of the LSS bi-spectrum in the folded case is much larger than that in the equilateral case \cite{Namikawa:2016b}. 

In order to see quantitatively how well the parameters $\xi_\lambda$ and $\xi_\kappa$ can be constrained, we follow Refs.~\cite{Takada:2004,Namikawa:2016b} and define the Fisher matrix:
\al{
	F_{ij} &= \sum_{\l_1\leq\l_2\leq\l_3}^{\lmax}f_{\rm sky} 
		\frac{B_{\l_1\l_2\l_3,i}B_{\l_1\l_2\l_3,j}}{\Delta_{\l_1\l_2\l_3}\mC{C}_{\l_1}\mC{C}_{\l_2}\mC{C}_{\l_3}}
	\,, \label{Eq:fisher}
}
where $B^\kappa_{\l_1\l_2\l_3,i}$\, is the derivative of the lensing bi-spectrum with respect to 
the $i$th parameter. 
$\Delta_{\l_1\l_2\l_3}$ is unity if all $\l_i$ are different, $2$ if two $\l_i$ are equal, 
and $6$ if all $\l_i$ are equal. 
The lensing power spectra, $\mC{C}_\l$, includes the reconstruction noise of the lensing measurement. 

The lensing reconstruction noise is computed by following the formula of Ref.~\cite{Smith:2010gu} 
which is motivated by the maximum-likelihood approach to reconstructing the lensing potential \cite{Hirata:2002jy,Hirata:2003ka}. The specifications of the CMB instrumental noise we use are summarized in Table \ref{Table:cmb}.

The constraining power at various scales is investigated by varying the maximum multipole $\lmax$ of the summation of \eq{Eq:fisher}. 
Derivatives are numerically computed based on the symmetric difference quotient. 
The fiducial values of the parameters are GR ones, i.e., $(\xi_\lambda,\xi_\kappa)=(0,0)$. 
The other cosmological parameters are fixed in the analysis because these parameters are severely 
constrained by other observables such as 
the primary CMB power spectrum and lensing potential power spectrum. 
The linear matter power spectrum is computed with {\tt CAMB} \cite{Lewis:1999bs}. 
We use the nonlinear correction to the matter power spectrum of Refs.~\cite{Smith:2002dz,Takahashi:2012em}. 

Fig.~\ref{fig:const:usig} shows the expected size of the statistical error on
the parameters $\xi_\lambda$ (left) and $\xi_\kappa$ (right) as a function of the maximum multipole considered, $\lmax$, when only one parameter at a time, either $\xi_\lambda$ or $\xi_\kappa$, is free to vary while keeping the other fixed. 
These plots show quantitatively how fast the constraints improve when smaller scales are included. Let us also note that the $1\sigma$ constraints do not vary simply as a power law of $\lmax$.\footnote{
In the case of the power spectrum, $C_\l$, if the derivative of $C_\l$
with respect to a parameter, $p$, is proportional to $C_\l$, the Fisher matrix for $p$ in the cosmic 
variance limit is simply given by 
$F_{pp} = \sum_{1\leq\l\leq\lmax}(\l+1/2) \simeq \lmax^2$ ($\lmax\gg1$). 
Then the $1\sigma$ constraint on $p$ varies as $1/\lmax$. 
Nevertheless, \eq{Eq:fisher} is not simply written as a function of $\lmax$ and 
the $\lmax$ dependence of $\sigma$ becomes more complicated. 
}
One also sees that the constraint on $\xi_\kappa$ is much better than that on $\xi_\lambda$. This is mostly due to the factor $2/7$ in front of $\lambda$; in addition, the terms proportional to $\lambda$ vanish for the folded configuration ($\l_1=2\l_2=2\l_3$ and its permutations) 
which has dominant contributions to the large-scale structure bi-spectrum \cite{Namikawa:2016b}. In any case, future CMB experiments will allow exploring a relevant part of the parameter space which is so far nearly unconstrained. 

Fig.~\ref{fig:const:2d} shows the error contours of the joint constraint on the parameters $\xi_\lambda$ and $\xi_\kappa$. 
The degeneracy between the two parameters is clearly seen. It does not change with $\lmax$, nor is it really broken by the non-linear corrections. 
This degeneracy comes from the first term of the kernel $F_2$ in \eq{Eq:MG:F2:loop}. 

Fig.~\ref{fig:const:sig} shows the dependence of the statistical error of $\xi_\lambda$ (left) and $\xi_\kappa$ (right) on $\lmax$, when the two parameters are simultaneously constrained. 
By comparing Fig.~\ref{fig:const:sig} with Fig.~\ref{fig:const:usig}, 
due to the parameter degeneracy, the $1\sigma$ uncertainty in each constrained parameter 
becomes much larger compared to that in the absence of the degeneracy. 

It is worth stressing that the correction due to the nonlinear growth beyond tree level 
significantly increases the total signal-to-noise of the bi-spectrum \cite{Namikawa:2016b}. Correspondingly, the constraints on the parameters become significantly tighter. 

\subsection{Impact of nonlinear loop correction on the lensing bi-spectrum}

\begin{figure*}
\bc
\includegraphics[width=89mm,clip]{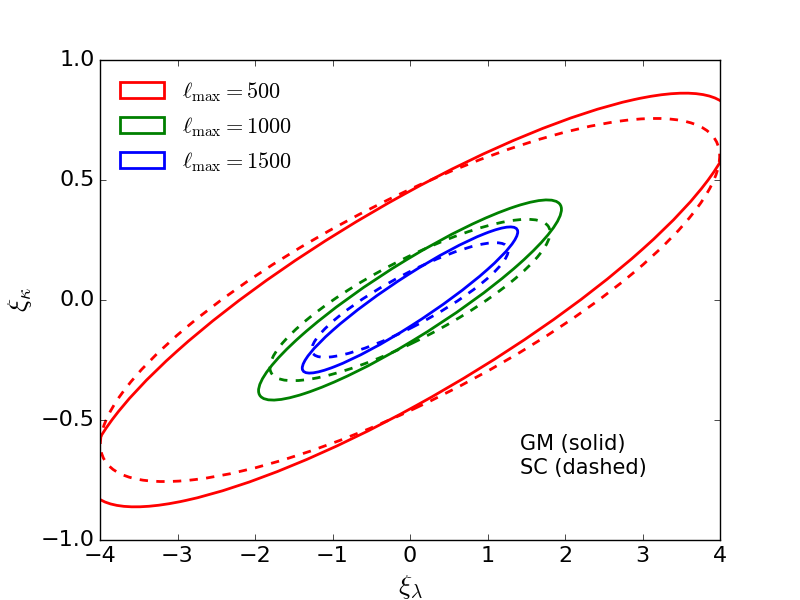}
\includegraphics[width=89mm,clip]{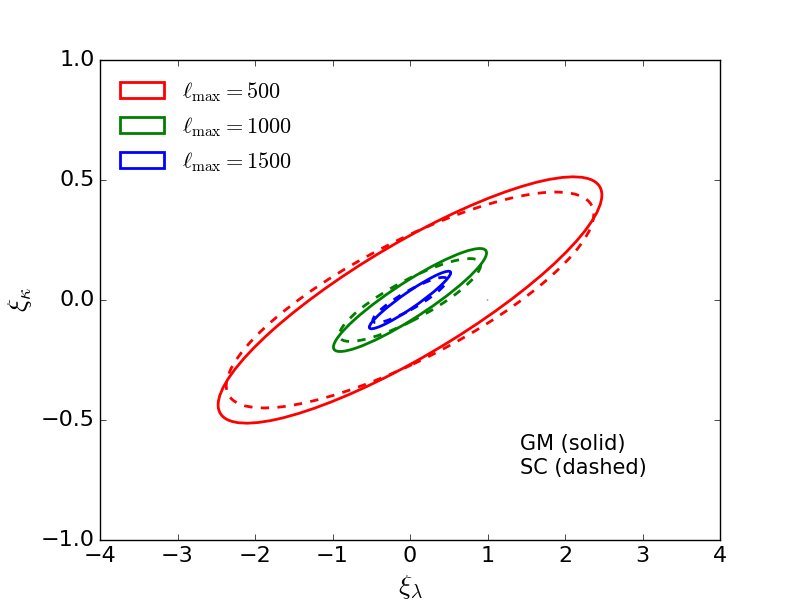}
\caption{
Expected constraints using the nonlinear loop correction of Ref.~\cite{Gil-Marin:2012} 
compared with that of Ref.~\cite{Scoccimarro:2001}. 
}
\label{fig:const:nonlin}
\ec
\end{figure*}

\begin{table}
\bc
\caption{
Expected bias in the parameters, $\xi_\lambda$ and $\xi_\kappa$, if the nonlinear loop correction 
of Ref.~\cite{Scoccimarro:2001} is inaccurate while that of Ref.~\cite{Gil-Marin:2012} 
correctly captures the loop correction. 
We also show between parentheses the expected $1\sigma$ constraints with the GM fitting parameters. 
}
\label{Table:bias} \vs{0.5}
\begin{tabular}{lccc} \hline \hline
 & $\lmax$ & $\Delta \xi_\lambda$ ($\sigma(\xi_\lambda)$) & $\Delta \xi_\kappa$ ($\sigma(\xi_\kappa)$) \\ \hline 
CMB-S4 &  500 & 0.26 (2.8) & 0.048 (0.57)  \\ 
       & 1000 & 0.45 (1.3) & 0.053 (0.28)  \\
       & 1500 & 0.49 (0.92) & 0.033 (0.20)  \\ \hline
CV     &  500 & 0.26 (1.6) & 0.048 (0.34)  \\ 
       & 1000 & 0.48 (0.65) & 0.053 (0.14)  \\
       & 1500 & 0.53 (0.35) & 0.026 (0.079)  \\ \hline
\end{tabular}
\ec
\end{table}

Our results imply that the study of the nonlinear growth in modified gravity theories is 
very important to best extract information in the lensing bi-spectrum in the era of CMB-S4 and beyond. 
We have used the fitting formula of GM for the matter bi-spectrum beyond the tree-level prediction (see \eq{Eq:GM}). Unfortunately, its validity  for deriving constraints on the modified gravity is not yet well studied. Indeed a complete treatment of nonlinear correction in modified gravity theories 
is highly involved and well beyond the scope of this paper. 
However, here, we get a feel of the sensitivity of our results to the specifics of the nonlinear correction by contrasting those obtained with GM with those derived with the SC fitting parameters (see \eq{Eq:SC}). 

The expected $1\sigma$ error contour in the two dimensional parameter space 
using the SC fitting parameters is shown in Fig.~\ref{fig:const:nonlin}. 
The expected constraints on $\xi_\lambda$ and $\xi_\kappa$ do not significantly deviate from those obtained 
in the GM case. This suggests than our results are actually quite robust against the details of the non-linear correction. 

Still, in addition to the effect on the uncertainties of the parameter constrains, 
the inaccuracy in the fitting formula could lead to a bias in the best fit parameters. 
Assuming that the GM fitting formula correctly captures the matter bi-spectrum 
beyond the tree level, we evaluate the expected bias in the parameter estimations 
by using on purpose an ``inaccurate" fitting formula as follows. Specifically, we regard the SC fitting formula as the inaccurate model, and compute the expected parameter bias as \cite{Joachimi:2009, Namikawa:2011a}
\al{
	b_i &= \sum_j \widetilde{F}^{-1}_{ij} \Delta_j
	\,, \label{Eq:bias}
}
where we define 
\al{
	\Delta_j &= \sum_{\l_1\leq\l_2\leq\l_3} f_{\rm sky} \frac{(B_{\l_1\l_2\l_3}- \widetilde{B}_{\l_1\l_2\l_3})\widetilde{B}_{\l_1\l_2\l_3,j}}{\Delta_{\l_1\l_2\l_3}\,\mC{C}_{\l_1}\mC{C}_{\l_2}\mC{C}_{\l_3}}
	\,, \label{Eq:Delta}
}
The quantities, $\widetilde{F}_{ij}$ and $\widetilde{B}_{\l_1\l_2\l_3}$, are the Fisher matrix and 
lensing bi-spectrum computed with the incorrect theoretical model of the 
fitting formula, respectively. 

Table \ref{Table:bias} shows the results of the parameter bias for several experimental 
specifications and maximum multipole. 
As the maximum multipole of the bi-spectrum increases, the bias also increases. 
In the CMB-S4 case, the bias is still well within the $1\sigma$ expected constraints. It will thus not require heavy theoretical work to solidify the constraints on modified gravity theories at loop level.
On the other hand, the bias for $\lmax\agt 1500$ in the CV limit is actually fairly significant 
compared to the $1\sigma$ constraint. 

\subsection{Impact of the screening mechanism on the lensing bi-spectrum} 
\label{subsec:screening}

At small scales, modified theories of gravity have to become close to GR in order to avoid violating the observational constraints based on solar system measurements. This is insured by a screening mechanism.
Following the treatment of Ref.~\cite{Fasiello:2017}, we model this screening mechanism through the following  
modification of the parameters in $F_2(\bk_m,\bk_n,z)$, 
\al{
	\lambda(z) & \to 1+(\lambda(z)-1)f(k_m,k_n,z) \,, \\
	\kappa(z) & \to 1+(\kappa(z)-1)f(k_m,k_n,z) \,,
}
where $(m,n)$ is $(1,2)$, $(2,3)$ or $(1,3)$ and
\al{
	f(k_m,k_n,z) = \exp\left[-\frac{k_m^2+k_n^2}{k_V^2(z)}\right]
    \,.
}
We further introduce a parameter, $v$, as $k_V(z)=v\,k_{\rm NL}(z)$ with $v\gg 1$. 
Choosing $v=10$, the constraints on $\xi_\lambda$ and $\xi_\kappa$ become weaker 
by at most $1\%$ ($10\%$) at $\lmax=500 \, (1500)$ 
compared to the case without the above screening effect. 
If $v=100$, the change is negligible. This shows that, as expected, our results are quasi-insensitive to the details of the screening mechanism which operates at much smaller scales than those probed by the lensing of the CMB. 

\subsection{Comparison with low-$z$ lensing bi-spectrum}

\begin{table}
\bc
\caption{
Same as Table~\ref{Table:bias} but for the lensing bi-spectrum probed with low-$z$ sources, in the cosmic-variance limit. 
}
\label{Table:bias:lowz} \vs{0.5}
\begin{tabular}{lccc} \hline \hline
 & $\lmax$ & $\Delta \xi_\lambda$ ($\sigma(\xi_\lambda)$) & $\Delta \xi_\kappa$ ($\sigma(\xi_\kappa)$) \\ \hline 
$z_s=1$ &  500 & 0.068 (0.11) & 0.027 (0.025)  \\ 
       & 1000 & 0.076 (0.044) & -0.030 (0.011)  \\
       & 1500 & 0.069 (0.027) & -0.074 (0.0068)  \\ \hline
$z_s=2$ &  500 & 0.18 (0.22) & 0.063 (0.049)  \\ 
       & 1000 & 0.24 (0.086) & 0.034 (0.021)  \\
       & 1500 & 0.27 (0.051) & -0.0081 (0.013)  \\ \hline
\end{tabular}
\ec
\end{table}

\begin{figure*}
\bc
\includegraphics[width=89mm,clip]{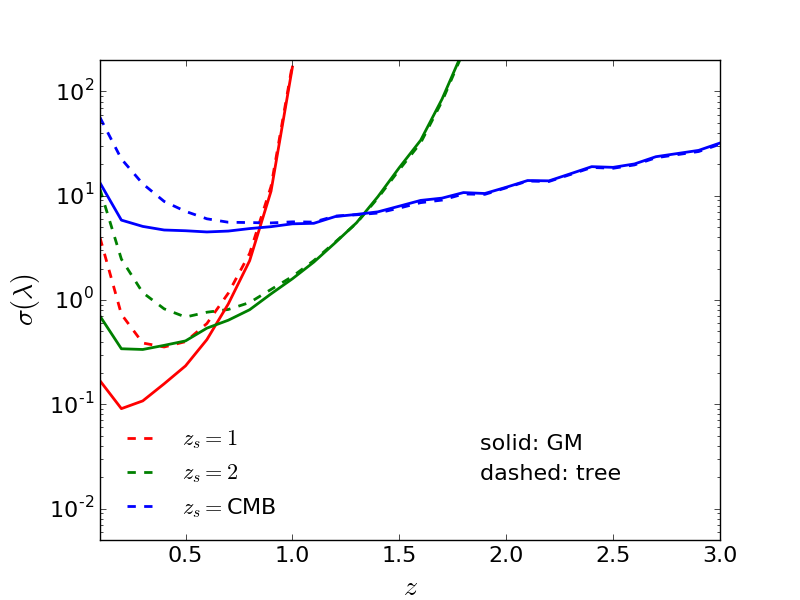}
\includegraphics[width=89mm,clip]{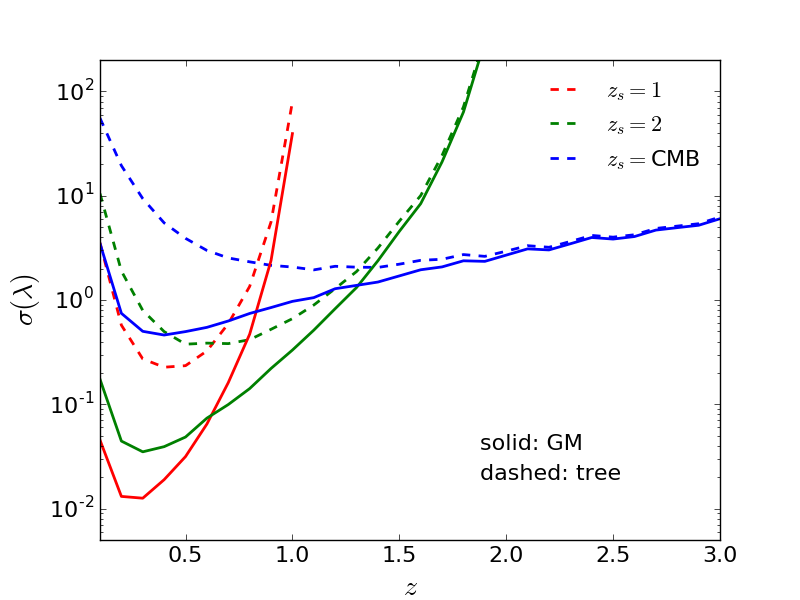}
\caption{
Expected $1\sigma$ uncertainty on $\lambda$ at each redshift (see \eq{Eq:lambda-z}) 
using multipoles up to $\lmax=500$ (Left) and $1500$ (Right) for various source redshifts, $z_s$. 
We do not include noise contributions in the covariance (i.e., this is the CV limit). Of course the impact of the non-linear correction beyond the tree level increases with decreasing redshift. 
}
\label{fig:const:lambdaz}
\ec
\end{figure*}

So far we have focused on the constraints using the CMB lensing bi-spectrum as a clean 
cosmological probe. Indeed the lensing bi-spectrum directly measures the underlying gravitational potential and,  unlike the galaxy bi-spectrum discussed by many previous works (e.g., \cite{Yamauchi:2017}), it
is immune to many observational and theoretical uncertainties, e.g., galaxy biases. 
The bi-spectrum of the cosmic shear is an alternative observational probe of modified gravity 
theories (e.g., \cite{Dinda:2018}) which is based on the same physics. The resulting constraints on the modified gravity theories are expected to be tighter than those obtained from the CMB lensing bi-spectrum (at least for monotonous deviations). 
However, the cosmic shear has significant uncertainties in, e.g., the theoretical modeling of the 
nonlinear growth of the large-scale structure, baryon physics, intrinsic alignment, 
and observational difficulties such as photo-z, PFS, and calibration. 

Table \ref{Table:bias:lowz} shows the expected bias and $1\sigma$ constraints on the parameters, 
$\xi_\lambda$ and $\xi_\kappa$, using the cosmic shear at $z_s=1$ and $2$, to be compared with the CMB results of Table \ref{Table:bias}. 
Compared to the latter case, the bias is larger than the $1\sigma$ constraint even at $\lmax=500$. 
The lensing bi-spectrum of the low-z galaxies is highly sensitive to 
the modeling of the nonlinear correction, and the constraints derived from the cosmic shear bi-spectrum 
would be easily biased by limitation of the accuracy of the nonlinear correction. 
In this respect, the constraints derived from the CMB lensing bi-spectrum 
will serve as an important cross-check of the results obtained from the cosmic shear bi-spectrum. We shall now see that CMB and shear lensing constraints are actually fairly complementary.  

Here we discuss which redshifts the lensing bi-spectra of CMB and galaxies are sensitive to. To that effect, 
we consider a time dependence of the parameter, $\lambda(z)$, varying according to a top-hat function selecting a redshift bin: 
\al{
	\lambda(z) = 
    	\begin{cases} 
        	1+\epsilon & (z_i\leq z\leq z_i+0.1) \\
            1 & (\text{otherwise})
    	\end{cases}
    \,. \label{Eq:lambda-z}
}
We compute the expected constraints on $\epsilon$ with varying $z_i$. 
Note that the results of the Fisher analysis do not depend on the fiducial value of $\epsilon$. 
We also change the source comoving distance, $\chi_*$, to include cases of galaxy weak lensing. 

Fig.~\ref{fig:const:lambdaz} shows the expected constraints on the parameter, $\lambda(z)$, 
at each redshift bin. We vary the maximum multipole, fitting function and source redshift. 
At high-$z$ ($z\agt 1$), the constraints from the CMB lensing bi-spectrum are much tighter than 
those from the lensing bi-spectrum of low-z sources. For both $\lmax$, the best constraints on the $z$-bin of $\lambda$ are obtained from $z_s=1$ at $z \alt 0.7$, from $z_s=2$ for $0.7 \alt z \alt 1.3$, and at higher $z$, the CMB constraints are the tightest. In addition, the plot indicates by how much the constraints depend on the somewhat uncertain non-linear correction beyond tree-level. 

While lensing analyses will initially be done independently for sources at various redshift, in the long run, the combination of measurements of the CMB and low-z lensing bi-spectra will allow a full blind reconstruction of the evolution of the modified gravity parameters. 

\section{Summary and discussion} \label{summary}

Besides their academic interest, modified gravity theories have been proposed as a means to interpret differently the observational evidences for Dark energy. Given the great success otherwise of Einstein theory at all scales, from the solar system tests of GR to the largest observational scales probed by CMB anisotropies, deviations have to be tenuous, and any search for such modifications are likely difficult and not immune to a host of troublesome systematics effects. 

Since the physics of CMB anisotropies is now well understood, the CMB lensing bi-spectrum will offer a very clean probe of modified gravity theories in the near future, compared to other cosmological probes such as galaxy clustering and optical weak lensing. In this paper, we quantitatively evaluated the expected constraints on a generic two-parameter model of the modified gravity theories assuming the specifications of near future and ultimate CMB experiments, which was not explored previously. 
This intermediate step allows avoiding a detailed comparison of the observation with a specific class of modified gravity models (in much the same way that the slope of the primordial curvature spectrum is a good agnostic point of contact between specific theories and observational constraints). 


We quantified the information coming from various scales and showed the impact of uncertainties in the theoretical description of the non-linear evolution of large scale structures. While ultimate, cosmic variance limited experiments will, in the long run, require further theoretical advances, we showed that our calculations are already adequate for the next decade experiments. And we further checked that details of the needed screening mechanism (to satisfy Solar system constraints) are at such very small scales that they have little impact at the scales which can be realistically probed with such CMB observations.    
 
Finally we point out the exciting prospect of blindly reconstructing the redshift evolution of the distortion parameters which we used to parametrize generically the beyond Horndeski class of modified gravity models (e.g., \cite{Kennedy:2018}). 
This will be achieved by performing a joint analysis of the bi-spectrum of CMB and source lensing, each providing the best constraints within different redshift range, in addition to allowing welcome cross-checks in the overlap region. 

\begin{acknowledgements}
TN and AT thank Daisuke Yamauchi for references and comments on modified gravity effects on the  lensing. This work was partly supported by the International Research Unit of Advanced Future Studies at Kyoto University. A.T. acknowledges financial support from MEXT/JSPS KAKENHI Grant Number JP15H05899 and JP16H03977. 
\end{acknowledgements}


\appendix

\bibliographystyle{mybst}
\bibliography{main,exp}

\end{document}